# Can Centauros or Chirons be the first observations of evaporating mini Black Holes?

A. Mironov[§]

*Theory Department, Lebedev Physics Institute and ITEP, Moscow, Russia*

A. Morozov[¶]

*ITEP, Moscow, Russia*

T.N. Tomaras[∥]

*Department of Physics and Institute of Plasma Physics, University of Crete, and Fo.R.T.H., Greece*

We argue that the signals expected from the evaporation of mini black holes - predicted in TeV-gravity models with large extra dimensions, and possibly produced in ultra high energy collisions in the atmosphere - are quite similar to the characteristics of the Centauro events, an old mystery of cosmic-ray physics.

## 1 Hypothesis

Among the various extensions of the Standard Model to energies beyond 1 TeV, one of the most attractive alternatives to the (Supersymmetric?) Great Desert Scenario is the TeV-gravity hypothesis with large extra dimensions [1]. According to it, matter particles and vector gauge bosons are open-string excitations, attached to a 3-brane (our world), which is embedded into compactified $D$-dimensional bulk space, where the closed-string excitations, including gravity, can propagate. This is the simplest possibility. Specific realizations of this idea and alternative scenaria may be found in [2]. Apart from a certain philosophic and aesthetic attraction of such models, they lead to the exciting possibility of experimental discovery of unification of the Standard Model with Quantum Gravity within the next few years, in the forthcoming accelerator, neutrino and cosmic-ray experiments [3, 4, 5]. Moreover, one could even claim that Quantum Gravity phenomena are already present in existing cosmic-ray data [6]. In the present paper we shall argue that the long-known Centauro-like events (CLEs) may be due to the formation and subsequent evaporation of mini black holes (MBHs), predicted in TeV-gravity models.

Our arguments are summarized in the comparative Table at the end, which describes (a) the characteristic properties of the observed CLEs, (b) the main features of the evaporating MBH events, and (c) the predictions of other popular theoretical models to explain the CLEs, namely in terms of fireballs, quark-gluon plasma and strange-matter physics. It is easy to use the Table to compare the positive and negative points of the various proposals. Most of the alternative explanations can qualitatively describe, often relying on fine tuning of relevant parameters, one or two peculiarities of the CLEs, but they either directly contradict other properties of the CLEs, or they take advantage of our incomplete knowledge of the physics of strong interaction processes and postulate features, which in some cases seem rather contrived. For instance, it seems to us that the hadronization/fragmentation part of various relevant processes, obtained by extrapolation of experimentally tested results to higher energies, may be more reliable than claimed in the context of some of these models. What is worse, the models involving strangelets, which is so far the most popular among the approaches to explain the CLEs, depend in addition on *ad hoc* assumptions about the properties of hypothetical strange matter, both with regards to its formation as well as its decay. As a rule, it is just optimistically assumed that these properties will exactly match those of the observed CLEs. At this stage, however, it is rather difficult to decide about the validity of these assumptions.

---

[§]E-mail: mironov@itep.ru; mironov@lpi.ac.ru
[¶]E-mail: morozov@itep.ru
[∥]E-mail: tomaras@physics.uoc.gr

The MBH hypothesis suffers from similar drawbacks, but, we believe, to a relatively lower degree. Among its advantages we would like to mention (a) its potential to explain all the CLE properties at once (unlike other approaches with assumptions targeted at these properties), (b) the fact that it offers somewhat less freedom to arbitrarily adjust different predictions, and (c) that it could easily be ruled out by already existing computer programs of Monte Carlo simulations, if applied to confront our hypothesis with the details of the Centauro characteristics. We believe that evaporating MBHs are strong candidates as the origin of unusual cosmic-ray events like the CLEs and the whole scenario deserves serious consideration and further analysis.

Section 2 is devoted to a brief presentation of the properties of MBHs in the context of TeV-gravity with large extra dimensions, while in Section 3 the main features of the Centauro events are described. A brief review of previous proposals to explain the origin of these events is the content of Section 4. Section 5 contains the presentation of the idea of evaporating MBHs as the origin of Centauros. We end with a few concluding remarks and the Table with the predictions of the MBH scenario and the comparison with the other proposals.

## 2 TeV-gravity models and mini Black Holes

Consider a flat 3-brane embedded into $D = 4 + n$-dimensional space-time with the topology $R^4 \times T_n$ of four-dimensional Minkowski space times an n-dimensional transverse space taken for simplicity to be a torus. Denote by $r_c$ the characteristic size of the compact space $T_n$. If $T_n$ is highly asymmetric, its largest dimensions play the main role in our considerations, so that $n$ is actually the number of the largest dimensions of comparable size. This means that $n$ can be smaller than six, even though the fundamental superstring theory is 10-dimensional. The strength of $D$-dimensional gravity in the bulk is defined by the fundamental Planck mass $M_P$, which expressed in terms of the string scale $M_s$ and coupling $g_s$ is $M_P^{D-2} = M_s^{D-2}/g_s^2$. In the framework of TeV-gravity models $M_P$ is of the order of a few TeV, or even less. Specifically, today's lower bound on $M_P$ from TeVatron data is $M_P \geq 0.6$ TeV [7], though in more sophisticated scenaria including warped dimensions it can be even lower. The strength of gravity at large distances on the brane (distances $\gg r_c$), which is defined by the physical Planck mass, $M_{Pl} \sim 10^{19} GeV = 10^{16} TeV$, is given in terms of $M_P$ and $r_c$ by $M_{Pl}^2 = M_P^{D-2} Vol(T_{D-4})$. Thus, $M_P r_c \sim \left(\frac{M_{Pl}}{M_P}\right)^{\frac{2}{D-4}} \sim 10^5$ for $D = 10$, i.e. $r_c \sim (10 MeV)^{-1}$.

A MBH [8, 9, 3] with mass $M_{BH}$, such that

$$M_P \ll M_{BH} \ll M_P \left(\frac{M_{Pl}}{M_P}\right)^{\frac{2(D-3)}{D-4}}$$

is essentially $D$-dimensional. As an interesting curiosity let us note that the upper limit, which is equivalent to the statement that the Schwarzschild radius $R_S \ll r_c$ and defines the boundary between $D$-dimensional and 4-dimensional black holes, is impressively large in TeV-gravity models. For instance, in $D = 10$ dimensions with $M_P \sim 1$ TeV, it is of the order of $10^{16 \cdot 7/3} M_P \sim 10^{37} TeV \sim 10^{13}$ kg.

The Schwarzchild radius of the MBH is

$$R_S = \frac{1}{M_P}\left(\frac{\pi}{2\kappa\Omega_{D-2}}\frac{M_{BH}}{M_P}\right)^{\frac{1}{D-3}}, \qquad (1)$$

while its Hawking temperature and entropy are

$$T_{BH} = \frac{D-3}{4\pi R_S} \qquad (2)$$

and

$$S_{BH} = \kappa\Omega_{D-2}(M_P R_S)^{D-2} = \frac{\pi}{2}M_{BH}R_S, \qquad (3)$$

respectively. In these formulas

$$\Omega_{D-2} = \frac{2\pi^{\frac{D-1}{2}}}{\Gamma\left(\frac{D-1}{2}\right)}$$

is the area of the (D-2)-dimensional sphere with the unit radius and $\kappa = (D-2)/32$. MBHs evaporate due to Hawking radiation with life time defined by the inverse Hawking temperature,

$$\tau_{BH} \sim \frac{M_P^{D-2}}{T_{BH}^{D-1}} \qquad (4)$$

For the purposes of the present paper we need the small MBHs with masses just slightly exceeding $M_P$. For large values of $D \sim 10$ the powers $1/(D-3)$ are small enough to make everything practically independent of the specific value of $M_{BH}$. In particular, such a MBH has a negligibly small life time $\sim 1/M_P \sim 10^{-27}$ s for $M_P \sim 1$ TeV. Most of the radiation is emitted in brane modes, since they are more numerous than the bulk ones [10]. The number of originally emitted particles is defined by the entropy

$$N_{in} \approx \frac{M_{BH}}{2T_{BH}} = \frac{4S_{BH}}{D-3} \qquad (5)$$

each of them carrying energy of the order of magnitude of $T_{BH}$. These primary particles can be anything with mass well below $M_P$, and emission of all degrees of freedom are equally probable (for instance, no $\alpha$-factors are present to suppress the creation of leptons). Hadronization of emerging quarks and gluons provides a lot of additional soft mesons with energies much lower than $T_{BH}$, and the actual multiplicity $N \gg N_{in}$. According to the stringy description of black holes, after evaporation they can leave remnants, which are some highly excited stringy states [11] (which, in particular, carry all the information, absorbed by the black hole) and further decay into light particles with the rate, regulated by the string coupling constant $g_s$ [12]. Black holes with masses below $M_P$ are fully quantum mechanical and is not clear if they can exist at all.

Finally, it will be assumed that the cross-section of MBH *production* is close to its geometrical cross-section,

$$\sigma_{BH}(E) = \eta\pi R_S^2(E), \qquad (6)$$

(where $R_S(E)$ is the Schwarzchild radius of a black hole with mass $E$), and does not depend on what kinds of particles collide to produce the MBH, or on whether they interact strongly, weakly, or electromagnetically). There is some controversy in the literature about this point [13] and the issue deserves further clarification. Nevertheless, we will assume that the existing arguments [14] [15] in support of the Thorne conjecture [16], i.e. eq. (6), are close to truth. The factor $\eta$ accounts for the radiation losses in the process of MBH formation. Existing simulations suggest that $\eta$ is definitely greater than 0.1.

# 3  The Centauro-like events

Centauro-like unusual events are observed in cosmic ray experiments high on the mountains Chacaltaya and Pamir [17, 18, 19], while no events of this kind have so far been observed at Mts. Kanbala and Fuji, where similar experiments started operating more recently. For a recent comprehensive rewiew see also [20]. The two storey detectors on Chacaltaya and Pamir consist of two parts, one above the other, separated by a target layer (polyethylene) and an air (or carbon) gap. Each part is a multiple sandwich of lead plates and sensitive layers (X-ray films and/or nuclear emulsion plates) to detect electron showers produced in the chamber through the cascade by the incident $\gamma$ or electron. Hadrons are detected since they produce a bundle of $\gamma$-rays through a nuclear collision with the lead. However, since the mean free path of hadrons is quite large (18.5 cm), most of the hadrons are registered in the lower chamber, while the electromagnetic component is registered exclusively in the upper chamber that is practically not transparent to it.

The usual interpretation of high energy cosmic ray events is as follows: an incoming cosmic particle (nucleon or nucleus) collides high in the atmosphere with a nucleus of the air. The collision is inelastic, the biggest fraction of energy being carried away by one or few "leading particles" (predominantly hadron jets, since creation of leptons is suppresed by powers of $\alpha = 1/137$), most of which have large ($p_T \sim E_{cm}/c \gg 10$ GeV/c) transverse momenta and are typically not observed in the detector. The detector captures a fraction $K \ll 1$ (gamma-inelasticity coefficient) of collision energy, in the form of numerous "soft" fragments with small $p_T \sim \Lambda_{QCD}/c < 1$ GeV/c produced during hadronization of the leading particles, remnants of a QCD string stretched between the scattered partons. These are the particles observed in the emulsion chambers and their energy can be measured by studying the track lengths in the emulsions and films. These particles are normally assumed to be $\pi$-mesons, $\pi^{\pm}$ being seen as hadrons in the chambers, while $\pi^0$s which "instantly" decay into two $\gamma$s, are seen as photons. Since isospin is conserved in strong interactions, the production of all three kinds of $\pi$-mesons is equally probable, this gives the normal ratio $\#photons/\#hadrons \equiv N_\gamma/N_h \sim 1$. In fact, the actual observed values of this ratio is closer to 2 and even more, since the electronic and photonic content increases through shower formation as the products of interaction descend towards the chamber. The total observed energy is a fraction of the initial energy $E_{lab}$ in the lab frame, $E_{obs} = KE_{lab}$. Assuming that the initial products were nucleons, the value of $K$ was estimated to be $\sim 0.2$, while if pions are produced initially, a higher value of $K \sim 0.4$ is preferred.

A whole new variety of events, which deviate from the above standard pattern, are observed in the energy region $E_{lab} \geq 500$ TeV. The most pronounced among them are the so called Centauro-like events (CLEs), whose characteristic properties are shown on lines 2-4 of the Table at the end. They have been distinguished into three groups, the Centauros, the mini Centauros and the Chirons [1], defined according to their hadron multiplicity, which is characteristically lower in mini Centauros and Chirons and the average transverse momentum $p_T$ of the hadrons in the event, which is much higher in Chirons. The main peculiarities of the CLEs are the following [20, 18]:

- The events observed take place close to the detector (from tens to hundreds of meters) rather than in the high atmosphere. This means that:

  (a) the incident initial cosmic-ray particle managed to penetrate deeply into the atmosphere,

  (b) immediate products of the collision are seen. Most likely the distance travelled by the initial fragments is not long enough for secondary showers to develop.

  (c) Similar events, if they occur at all in high atmosphere, are either not detected by the chambers or do not look anomalous.

- The events are hadron-rich, i.e. the ratio of energies of the observed photons and hadrons is unexpectedly low (for instance, in the single superclean Centauro II event of Chacaltaya it is just zero), which means that there are very few $\pi^0$s created. Among the 600 total number of highly energetic events registered in the Chacaltaya and Pamir joint chambers [21], the Centauro-like (or hadron-rich) constitute up to 7%, or even up to 10%. However, this figure would be as low as $\sim 0.1\%$ if one was to take into account only the super clean Centauros (see footnote 10).

---
[1]This is the name used in the literature, even though it is rather inappropriate for a whole class of Centauro events. According to Greek mythology, there were many Centauros (creatures half-men half-horses), but only one of them was called Chiron ().

- Detected hadrons have average $p_T \geq 1.5$–$15$ GeV/c, much larger than usually observed. As mentioned above, in the usual cosmic-ray showers the mesons produced have $p_T < 1$ GeV/c.

- Sometimes, traces are left by some strongly penetrating cascades (the so-called mini-clusters, i.e. clusters of particles with very small lateral spread corresponding to very low mutual transverse momenta $\sim 10$–$20$ MeV/c) travelling at ultrasmall angles ($\leq 10^{-6}$ radians) in the forward direction of the family, interacting abnormally weekly with the emulsion and, hence, leaving traces of unusual type (anomalous transition curves) called halo. They cannot be interpreted by standard means and are proposed to be non-electromagnetic remnants of an incoming cosmic particle, continuing to move along its original trajectory. The halo is usually observed in the Chiron events, which are characterized by especially large transverse momenta ($\sim 10$–$15$ GeV/c), smaller multiplicity ($10-20$ particles as compared to $60-90$ of the normal Centauro events) and existence of unusual hadronic component with the short mean free path $\sim 1/3$–$1/2$ of the nucleon geometrical collision mean free path.

- All these anomalous effects appear only for high enough energies of incoming particle, while at low energies there is nothing similar, i.e. the effect has an energy threshold (for figures, see the Table; these figures depend on various simpifying assumptions and rough estimates, as for instance of the value of $K$).

## 4  Previous proposals to explain the Centauro-like events

The two main lines of thought to explain the CLEs attribute their anomalous features either to a conjectured change of the properties of strong interactions at high energies, or to exotic cosmic particles. As it will become clear in the next section, our **hypothesis** is sort of a combination of these two approaches.

**Exotic incoming particles.** *Exotics* for cosmic ray physics are all stable particles but nucleons, nuclei and photons. In particular, the leptons are exotics. The only Standard Model particle of this type, which is really a component of the cosmic rays, is the neutrino. Electrons, in particular, which are light and charged cannot appear in high-energy cosmic rays. A sufficiently high flux of cosmic neutrinos with energies in the kTeV range is expected to exist [22, 23]. They do not participate in strong and electromagnetic interactions, while their weak interactions are no longer negligible at energies above $M_W$, the typical cross section being

$$\sim \frac{\alpha^2}{M_W^2 \sin^4\theta_W} \sim \frac{0.1}{TeV^2} \sim 10^{-38} m^2. \tag{7}$$

Remarkably, with this cross-section the mean free path of high-energy neutrinos inside the Earth is comparable with the radius of the Earth $\lambda \sim (10^{-10}m)^3/10^{-37}m^2 \sim 10^7 m$. Formula (7) is to replace the usual

$$\frac{\alpha^2 s}{M_W^4 \sin^4\theta_W} \sim \frac{2\alpha^2 m_p E_{lab}}{M_W^4 \sin^4\theta_W} \sim 10^{-38} \frac{E_{lab}}{TeV} m^2,$$

valid for $s \ll M_W^2$ (i.e. for $E_{lab} \ll 10$ TeV). For better estimates of the cross-section, taking into account the quark-distribution functions one should see [24]. There the authors conclude that the actual cross-section continues to grow slowly with energy as $E_{\text{lab}}^{0.363}$, at least up to energies $E_{\text{lab}} \sim 10^8$ TeV.

**Exotic globs.** This approach just postulates the properties of hypothetical particles that are not Standard Model particles but some "extraterrestrial objects" that give rise to the CLEs. The required properties are:

- The hypothetical particle has a non-vanishing mass $M$ of the right value to serve as a threshold for the energy of its decay events.

- It can either decay instantly or interact with atmospheric nucleons, but it anyway should live in the atmosphere long enough to penetrate deeply and reach the detectors before its decay/collision.

- The products of reaction (decay or collision) do not contain (or contain few) $\pi^0$s and produce other hadrons and/or leptons with somewhat transverse momentum $p_T$ higher than expected in QCD.

- If necessary, after the decay, the object may leave non-electromagnetic, strongly interacting projectiles.

Various kinds of hypothetical quark matter have been considered as concrete proposals for the role of exotic globs. Today the most popular candidates to play the role of such extraterrestrial objects are the cosmic strangelets [25, 26].

**Strangelets.** If an $s$-quark is embedded into the Fermi liquid of $u$ and $d$ quarks, it cannot decay because all the energy levels for the would-be created $u$ or $d$ quarks are occupied. At the same time, it can be energetically profitable for some $u$ and $d$ quarks to turn into $s$-quarks since this could lower the level of the Fermi surface. In this way, one could presumably form some stable "strange matter" with non-vanishing concentration of strange quarks [27]. Ordinary matter consisting only of $u$ and $d$ quarks is known to be less stable than the nuclei with the same quantum numbers (just because these latter do not decay into quark matter) and can only be artificially created in the form of unstable quark-gluon plasma (QGP). However, one could expect a stable configuration to exist, formed from roughly equal numbers of quarks of all three kinds (so that the charge/mass ratio is small). It is believed that "strange stars" and "primordial strange matter" may exist in the Universe. Furthermore, it has been argued [28] that they can emit "strangelets", relatively small pieces of strange matter, which will then be among the components of cosmic rays bombarding the Earth. Alternatively, strangelets could be created in nucleus-nucleus collisions in high atmosphere. Whatever their origin, strangelets while travelling through the atmosphere slowly loose mass, until they reach the phase transition point and instantly decay. This idea is considered as the most popular explanation of the CLEs, because the pattern of strangelet decay should be different from that of an ordinary hadronic collision. Even though it is not very clear how the remaining *concrete* peculiarities of CLEs will be reproduced in this way, it is instructive to describe briefly the rather sophisticated reasoning of the strangelet scenario to explain the smallness of the ratio $N_\gamma/N_h$. The small value of this ratio is qualitatively ascribed to the dominance of strange particles in strangelet decay products. This happens not only because the strange quarks are present and $u$ and $d$ can combine with them in forming mesons or baryons, but also because the formation of bound states of $u$'s and $d$'s is damped inside the Fermi liquid. According to the most radical version of this scenario [25], immediate products of the decay are only baryons, while $\pi$'s are created only in decays of strange baryons, with at most one $\pi$ per baryon. Thus, the fraction of $\pi^0$s is about $\sim 1/3$ times $\sim 1/2$ for the probability to have strange baryons created. Moreover, not all but rather 30-50% of these $\pi^0$s have sufficient energy to produce $\gamma$-quanta, observed in the detectors, since for this the energy/quantum should exceed a few TeV, for x-ray films depending on experimental conditions, or 0.1 TeV for nuclear emulsion plates. Altogether this leads to an estimate as low as 5% for the fraction of relevant $\pi^0$s (and thus 10% of $\gamma$'s) as compared to the total number of emitted baryons, registered as hadrons in the detector. Given that each $\pi^0$ produces two photons one ends up with the figure $N_\gamma/N_h \sim 0.1$ for the ratio of photons to hadrons in the strangelet decay event.

**Peculiarities of strong interaction at high energies and densities.** In the absence of a satisfactory solution of the confinement problem, the implications of strong interactions in any experimentally unreachable domain are not reliably understood. Thus, the properties of the hadronization/fragmentation process at ultra-high energies, the precise nature of the hypothetical diffractive hadronic fireballs [29], the behaviour both of the quark gluon plasma (QGP) with disoriented chiral condensate (DCC) [30] and, as mentioned above, of strange matter [31], used in one way or another in all existing proposals to explain the Centauro events, are all subject to speculation.

Having presented the strangelet idea about the origin of the CLEs, let us now briefly review how the small $N_\gamma/N_h$ ratio is dealt with in the context of the, so called, DCC approach [30, 32]. According to it, the starting point of the CLEs is the formation of hot QGP as a result of high energy collisions of cosmic ray hadrons or heavy ions with nucleons or nuclei of the atmosphere. In this initial state the quarks and the gluons are not confined to hadrons and the chiral symmetry is restored. In the language of the linear $\sigma$-model for the description of the chiral transition in QCD, the field $\Phi = (\sigma, \pi^i)$ whose components are the isosinglet $\sigma$ and isotriplet pion effective fields, starts at the symmetric value $<\Phi>= 0$. Subsequently, as the hot QGP expands and cools, the chiral symmetry breaks spontaneously and a condensate with non-vanishing $<\Phi>$ will be formed. But, in the begining $<\Phi>$ will not necessarily be oriented along the $\sigma$-axis, corresponding to the true QCD vacuum at zero temperature. It may lie, instead, in the $\pi^+\pi^-$ plane (orthogonal to $\pi^0$). In this case, during its relaxation to the ordinary QCD vacuum ($<\sigma>\neq 0, <\pi^i>= 0$), only charged pions will be created, and a small ratio of $N_\gamma/N_h$ will be observed. Correspondingly, mostly $\pi^0$s will be created if, instead, the condensate points initially along the $\pi^0$ direction. This will lead to hadron-poor events, with anomalously high $N_\gamma/N_h$. Such anti-Centauro events have indeed been observed but in much smaller numbers in a very different baloon experiment JACEE [33] at the top of the atmosphere, which, on the other hand, has not yet observed any CLE. The fact that a lot fewer anti-Centauro compared to Centauro events have been observed is ascribed to the lower probability for the chiral condensate to point along the $\pi^0$ *axis* than along the $\pi^+\pi^-$ *plane*, while the non observation of ordinary CLEs at JACEE may be due just to the small area and short time of exposure in the JACEE experiment, as compared to the Chacaltaya-Pamir ones. In fact, it is generally quite ambiguous to compare these experiments because of different conditions. Overall, the DCC is considered an elegant theoretical idea to explain the low level of $\pi^0$ in the CLEs. However, even if the DCC exists at all, it does not seem sufficient to explain the other CLE characteristics.

A quick glance at the Table in the end, which summarizes the predictions of the above most succesful so far proposals, convinces the reader about the common belief that none of them can account for all the peculiarities of the Centauro events (see, e.g., [19]).

## 5   Evaporating MBHs as an explanation of the CLEs.

Very briefly, according to this hypothesis one envisages MBHs formed by the collision of ultra high energy cosmic ray particles with nucleons or nuclei of the atmosphere. The MBH almost instantly evaporates to give a number of Standard Model particles. One of these initial fragments evolves into a shower, with the characteristics of the CLEs, recorded in the detectors.

The way and extent to which such a scenario can account for the origin of the CLEs is as follows:

- *Energy threshold.* First, note that the MBHs can only be created at energies $E_{cm} \geq M_P$, i.e. the above phenomenon will indeed be characterized by an energy threshold.

- *Production.* The cross section of MBH production in hadronic collisions is $\sim \frac{1}{M_P^2} = \frac{1}{m_p^2}\left(\frac{m_p}{M_P}\right)^2$, i.e. $\sim 10^{-6}$ of the typical hadronic cross section. This ratio is big enough to guarantee the observation of evaporating MBHs in the high-luminosity accelerator experiments in TeVatron and LHC, if the TeV-Gravity models are true at all and if the accelerator center-of-mass energy exceeds $M_P$. However, it seems too small to explain not only the optimistic 20% level of the hadron-rich events, but even the very modest level $10^{-3}$ of the superclean Centauro events as compared to the ordinary hadronic production cross section in high-energy cosmic rays. This is the main problem for our hypothesis and seems to rule out the possibility that Centauros are decays of MBHs produced in *hadronic* collisions, say between cosmic and atmospheric nucleons.

A possible way out is to assume that the MBHs are created in collisions of more exotic cosmic ray particles. For instance, one could rely on the by-now standard expectation that MBHs in cosmic rays can be produced by cosmic neutrinos colliding with atmospheric particles [4]. Then, the frequency of such events will depend on the cosmic neutrino flux, which is not yet directly measured and can be taken only from astrophysical models and estimates [23, 22]. The usual expectation is that the flux is sufficient to make the MBH events observable, provided $M_P$ is in the TeV region. A possible neutrino-MBH origin of CLE would also explain their deep-penetration property, i.e. the fact that events are observed deep in the atmosphere close to the detectors. The cross section of MBH production by neutrino-nucleon collision is $\sim R_S^2 \sim 1/TeV^2$, i.e. comparable with (7). Thus, in contrast to the nucleon-nucleon case, the MBH creation is in no way damped as compared to the Standard Model interactions, and sufficiently high neutrino flux could make the number of MBH events comparable with that of nucleon-nucleon collisions, without creating a lot of new cosmic ray events due to conventional neutrino-nucleon scattering.

It is straightforward to obtain a rough estimate of the number of CLEs based on current figures for the neutrino flux $\Phi_\nu$. Since we are interested in neutrino energies of order $10^6 - 10^7$ GeV in the lab frame, we can use the estimates for the gamma-ray burst muon neutrino flux given in [22, 34]. Their analysis leads to $\sim 20$ neutrino-induced muon events in a cubic-km water or ice detector per year. Since the cross-section of MBH production by neutrinos is of the same order of magnitude as that of muon production, we would expect approximately the same number of MBH production events for each kind of neutrino. Of course, one also has to multiply these figures by the number of jets (particles) originally emitted from the MBH (of order 10-20 according to (5)), which ultimately leads to $\mathcal{O}(10^3 - 10^4)$ events. However, since the CLEs we are interested in are produced in the atmosphere as opposed to water or ice, taking into account the density of the atmosphere and its depth, we expect for the CLEs the intensity of $10 \div 10^2$ km$^{-2}$ year$^{-1}$, which is much less than the claimed intensity $10^3 \div 10^4$ km$^{-2}$ year$^{-1}$ of Centauro events at the Chacaltaya altitude [20, 35]. Note, however, that the estimates of the neutrino flux in [22, 34] can be easily enlarged by $1 \div 2$ orders of magnitude (see [36] and the $\gamma$-ray bound there) giving rise to a CLE rate consistent with observation.

A possible objection to the idea of $\nu$-produced MBHs is related to the isotropic distribution of incoming neutrinos, which should result in roughly isotropic distribution of the observed showers. This is in contrast to the case of ordinary cosmic rays, for which the Earth is effectively 36 times thicker in the horizontal direction, and, consequently, horizontal showers are practically excluded (see [37] on special searches for such showers). Moreover, the upward directed showers caused by neutrino collisions with Earth nuclei would, a priori, also be possible. Finally, there should not be any dependence of the probability of creating MBHs on the altitude above sea level. The experimental situation in connection with these issues is unclear: (a) No up-side-down showers were observed, but we are not aware if such attempts have been made at all; this is basically determined by the geometry of the experiment. For example, the upside-down showers may be fully absorbed by the basement. In addition, given the fact that the mean free path of these ultra highly energetic neutrinos is comparable to the earth radius, their flux coming through the Earth will be somewhat diminished. (b) Fewer Centauros in Pamir, 4300 m (600 $g/cm^2$) than in Chacaltaya, 5200 m (540 $g/cm^2$) are observed so far. This difference in the numbers of the Chacaltaya and Pamir events has been considered in [26] as a confirmation of the atmospheric strangelet model. However, we do not think this data can in any way be decisive, especially given the fact that on the even higher Mts. Kanbala (5500 m or 520 $g/cm^2$) no Centauros at all have been reported yet.

Still, it should be pointed out, that even if cosmic neutrinos turn out to be inadequate for efficient MBH production, one could still envisage other possible sources of MBHs, namely some other exotic cosmic particles, presumably not involved in strong interactions, or some yet unknown component of dark matter. In fact it is possible that even some heavy, stable, strongly interacting particles, including even some kind of strangelets, can interact with the atmospheric nuclei rather weakly, so that – like in the case of neutrinos – their interaction through creation of MBHs can become competitive.

- *Evaporation.* The MBHs evaporate instantly. For $M_P \sim 1$ TeV their life time in the rest frame is $\sim 1/M_P \sim 10^{-27}$ s. They produce with equal probabilities particles of all kinds, at least from among the $\sim 120$ of the Standard Model, with spin and color degrees of freedom counted independently. The original multiplicity of the decay products is $\sim S_{BH} = \frac{\pi}{2} M_{BH} R_S$, while their characteristic energy and $p_T \sim T_{BH} \sim M_P \sim 1$ TeV. These are presumably the leading particles and in the centre-of-mass frame of the MBH they are emitted isotropically. This looks like a typical fireball situation, but the fireball (MBH) temperature is very high $\sim T_{BH} \sim 1$ TeV and only a fraction of radiated particles, namely 1-2 of the initially produced, is observed in the chambers. Therefore, the properties of the soft component, its $p_T$-distribution, the level of $\pi^0(2\gamma)$ and the coefficient $K$ (in this case, it measures the ratio of the observed energy to the total mass of the black hole) have no reason a priori to be close to the standard *hadronic* quantities. Moreover, since among the initially produced particles the fraction of $u$ and $d$ quarks is very small, the argument of the strangelet scenario can be applied even more succesfully to explain the low value of $N_\gamma/N_h$.

   Alternatively, one may argue along the lines of the DCC approach described above and which is considered quite plausible in explaining the hadron richness of the CLEs. It is reasonable to expect that the MBH after its evaporation and the subsequent expansion, cooling and production of secondary radiation, mostly gluons, will end up in a hot QGP state of temperature $T \sim 100$ GeV. From this point on, its evolution will follow the DCC picture, only at higher temperature and, consequently, hadron richer final products, than in the original proposal.

   Existing simulations for the search of MBHs in accelerators [5] clearly imply that the $p_T$ distribution is broader and the ratio photons/hadrons lower than for ordinary hadronic interactions. Besides, the neutrino-produced MBHs are predicted to give rise to deep penetrating showers. If these results match the CLEs not only qualitatively but also *quantitatively* is still to be checked.

- After evaporation, the MBH is expected to produce an excited stringy state, which, in principle, will decay into light (compared to $M_P$) particles. Their rate of decay depends strongly on the excitation level, and their life time can be considerably larger than $\tau_{BH} \sim M_P^{-1}$ [12], so that these stringy states could probably play the role of the projectiles with unusual behaviour, potentially responsible for the halos.

- At least three different families of CLEs with rather different characteristics seem to be observed (see Table). If they are indeed well separated from each other, rather than forming a smooth distribution, they might be attributed to decays of different types of MBHs or those of their remnants. Alternatively, one could attribute, say, Centauros to decays of strangelets, while Chirons to evaporation of MBHs. One should also check whether other unusual events, such as the ones observed in primary cosmic rays at high atmosphere in experiments with the baloon [38] (objects observed with large masses ($> 0.35$ TeV and even $> 1$ TeV) and relatively low charges (15-45)), which are usually interpreted as strangelets, could actually be charged black holes.

# 6  Conclusion

A preliminary analysis was presented of the idea that the well-known Centauro or even better the Chiron events are due to evaporating mini black holes with mass of $\mathcal{O}(\text{TeV})$, as predicted to exist in the TeV-gravity models with large extra dimensions. A comparison with other proposals to explain these events was made. The present experimental data on the main characteristics of the Centauro and/or Chiron events do not seem to rule out the hereby proposed explanation. In fact, a quick glance at the Table indicates that the mini black hole proposal looks more promising than other ideas put forward in the literature.

Our study, however, is just a first step towards a thorough analysis of several unusual events in cosmic ray physics, and of the possibility that they too may be attributed to mini black holes. A lot more has to be done. For instance, a more detailed numerical analysis and simulation of the production and decay of a MBH, such as the ones performed for LHC is crucial and might easily rule out the present interpretation. Also, the study of the evaporation of rotating and/or charged mini black holes may open new directions of theoretical explanations of some of the unusual events. In the experimental front, it would be very important to obtain more information and better statistics on the properties of these events, as well as to investigate the directional dependence of the events, their altitude distribution and so on. All this is crucial information which distinguishes the various proposals.

There is considerable scepticism concerning the experimental evidence itself for the existence of the, so called, Centauro interaction. The main reason are the unsuccessful attempts to produce Centauros in collider experiments [39]. The fact that all these efforts concentrated on the study of the central region of the rapidity space may, in view of the above scenario, be the reason behind their negative results. On the other hand, the remarkable properties attributed to the reported events, have excited the imagination of theorists and experimentalists alike. In addition, it is so important in our view to find out whether the fundamental quantum gravity scale is as low as a few TeV, as suggested by the type-I superstring theory models with large extra dimensions, that all possible consequences deserve exhaustive theoretical and experimental study. It is in this spirit that we analysed and presented the above model for the origin of the Centauro-like events.

## Acknowledgements

We would like to acknowledge helpful discussions and correspondence with D. Goulianos, J. Iliopoulos, G. Landsberg, A. Leonidov, T. Sjostrand and P. Sphicas. This work was supported in part by the EU under the RTN contracts HPRN-CT-2000-00122 and HPRN-CT-2000-00131. The A.M.s acknowledge the support of two NATO travel grants and the hospitality of the Department of Physics of the University of Crete, where this work was done. This work was also partially supported by the Federal Program of the Russian Ministry of Industry, Science and Technology No 40.052.1.1.1112 and by Volkswagen Stiftung, by the grants INTAS 00-334, RFBR 01-02-17682a (Mironov), INTAS 00-561, RFBR 01-02-17488 (Morozov) and by the Grant of Support for the Scientific Schools 96-15-96798 (Mironov).

Table 2

|  | Number of free parameters | Deep pen.[3] | Threshold Energy[4] | $N_{CLE}/N_{total}$ [5] | Mass of the IS[6] | Temp. of the IS | Life/decay time of the IS[7] | Multiplicity of hadrons | $N_\gamma/N_h$ | $p_T$-distribution | Halo |
|---|---|---|---|---|---|---|---|---|---|---|---|
| Usual events[8] |  | - | No |  |  |  |  | $\sim 35$ | $\sim 2$ | 0.4-0.8 GeV/c | - |
| Centauros |  | + | 1.8 (6.9) TeV[9] | 2.5%[10] |  |  |  | 60-90 | low[11] | 1.7 GeV/c |  |
| Mini Centauros |  | + | 1.3 TeV | 5%[12] |  |  |  | 10-20 | low[11] | 1.7 GeV/c |  |
| Chirons |  | + | 1.8 TeV | 7%[13] |  |  |  | 10-20 | low[11] | 10-15 GeV/c | + |
| MBH | **few**: $M_{BH}, M_P$ [14] | + | $M_{BH} > M_P$ | $\sim 0.0001\%$[15] **few % ?**[16] | $M_{BH}$ | $\sim M_P$ | $10^{-27}$ s [17] | $N_{in} = \frac{M_{BH}}{2M_P}$ $N_{obs} = ?$[18] | **low** | **very wide and flatten**[19] | + |
| Fireballs | model-dependent | - | $2 \pm 0.5$ TeV |  | 180 GeV | $T_{QGP}$ | $10^{-23}$-$10^{-24}$ s | **15 to 100**[20] | $\simeq 2$ | **wide** | - |
| QGP/DCC | DCC direction | - | $E \gg T_{QGP}$[21] |  |  | >150 MeV | $10^{-23}$-$10^{-24}$ s |  | **low**[22] | **narrow**[23] | - |
| Strangelets | $s$-quark density baryon number | + |  |  |  |  |  |  | low[24] |  |  |
| [31] | many | + | $2 \pm 0.5$ TeV |  | 180 GeV | 130 MeV | $10^{-23}$-$10^{-24}$ s [25] | **75** | low | **wide** | + |
| [26] | many | + | 300-400 GeV | Depends on flux | 300-400 GeV |  |  |  | low | narrow | + |

---

items are underlined. Minuses also imply disagreement.

[3] *Deep penetration* is a very important feature of the events observed, since all the events discussed have happened very close to the detectors as compared with the typical distance.

[4] All the energies in experimentally observed events are calculated assuming for the inelasticity coefficient $K = 0.2$. This means that the mass of any intermediate state (IS) is 5 times smaller than this energy.

[5] $N_{total}$ here is the number of *all* events with energies greater than 1 TeV.

[6] This is the typical mass of the IS relevant to the particular model. The emergence of the IS in some cases is justified by the corresponding scenario, in others it is just assumed.

[7] If the IS lives long enough, its main time characteristic is how fast it decays.

[8] The figures for multiplicity and $p_T$-distribution in this line are obtained by an extrapolation of experimental data to the Centauro energies.

[9] This number refers to scattering of the cosmic particle on a nucleon (nucleus with mass $14 m_{proton}$) in the atmosphere. Despite of the fact that in cosmic rays one measures energies in the laboratory frame, from now on and for convenience the energies reported will be the in center of mass system and for scattering on a nucleon. In principle, the energies of the CLEs go up to 4 TeV [20], however, these are usually not pure Centauros, and including them into the Centauro family would mean that up to 20% of all families with total energy larger than 1 TeV should be recognized as members of the Centauro species.

[10] If one takes into account only the super clean Centauro events the estimated ratio becomes $> 10^{-3}$ [19]. However, here is quoted a more "balanced approach", which gives 7 events among 305 families with energies larger than 1 TeV, observed so far in the Chacaltaya and Pamir joint chambers [21].

[11] In accordance with [20], the number of $\gamma$-quanta observed is either just 0 (for the so-called super clean Centauros) or so small that one may easily ignore the electromagnetic component of the event. However, these figures depend on how optimistic is the treatment of the experimental data. In any case, their number is less than the number of hadrons. Note that one would normally expect the hadronic component to constitute $\leq 30\%$ of the total visible energy.

[12] 15 of 305, see footnote 10.

[13] 21 of 305, see footnote 10.

[14] One could also add the charge and the angular momentum, but here we consider only Schwarzchild black holes.

[15] In the hadronic channel.

[16] In the $\nu$+nucleon channel. This figure of a few percent is the expected rough estimate [23, 22], though it depends on the neutrino flux; see also the discussion in the text.

[17] This estimation assumes just complete decay of the black hole to radiation. However, as mentioned in the text, it may produce remnants, which live much longer [12, 40].

[18] The multiplicity of the particles observed $N_{obs}$ should be calculated using fragmentation functions and strongly depends on the BH parameters. Besides, it is strongly influenced by the particles missed, which also depends on the details. Still, one should at least expect that $N_{obs}$ is much larger than $N$.

[19] Concrete figures strongly depend on the BH parameters. For instance, estimations in [5] for MBHs with masses between 12 and 14 TeV (the estimation being done for LHC is for energies much larger than those relevant to the Centauro events) and $M_P = 1$ TeV give typical $p_T \sim 25$ GeV/c, the MBH signal being much bigger than the QCD one already starting from around 50 GeV/c.

[20] Multiplicity $\sim 15$ is predicted for the mini-Centauro events, while the multiplicity $\sim 100$ corresponds to Centauro events.

[21] The threshold energy is determined by the condition necessary for the QGP to emerge, i.e. by thermalization conditions. However, this is in no way a threshold effect.

[22] See the discussion in the text about DCC and QGP.

[23] See, however, [41].

[24] See the discussion of strangelets in the text.

[25] This is the typical decay time of a strangelet that is created high in the atmosphere and has to travel 15 km before it decays. Its lifetime is of the order of $10^{-9}$ s [31].

[41] J.Bjorken, SLAC-PUB-6488 (1994)